\documentclass[prl,aps,reprint,floatfix,showpacs,twocolumn,superscriptaddress]{revtex4-1}

\usepackage{graphics}
\usepackage[dvips]{graphicx}
\usepackage{color}
\usepackage{subfigure}
\usepackage{url}
\usepackage[linktocpage,colorlinks=true,linkcolor=blue,citecolor=blue]{hyperref} 

\begin{document}





\title{Flow-induced nonequilibrium self-assembly in suspensions of stiff, apolar, active filaments}

\author{Ankita Pandey}
\affiliation{Indian Institute of Technology, Madras, Chennai 600036, India}
\author{P. B. Sunil Kumar}
\affiliation{Indian Institute of Technology, Madras, Chennai 600036, India}
\author{R. Adhikari}
\affiliation{The Institute of Mathematical Sciences, CIT Campus, Chennai 600113, India}


\date{\today}


\begin{abstract}

Active bodies in viscous fluids interact hydrodynamically through self-generated flows. Here we study spontaneous aggregation induced by hydrodynamic flow in a suspension of stiff, apolar, active filaments.  Lateral hydrodynamic attractions in extensile filaments lead, independent of volume fraction, to anisotropic aggregates which translate and rotate ballistically. Lateral hydrodynamic repulsion in contractile filaments lead, with increasing volume fractions, to microstructured states of asters, clusters, and incipient gels where, in each case, filament motion is diffusive. Our results demonstrate that the interplay of active hydrodynamic flows and anisotropic excluded volume interactions provides a generic nonequilibrium mechanism for hierarchical self-assembly of active soft matter.

\end{abstract}

\pacs{}

\maketitle

Filamentous structures along which chemical energy is converted to mechanical
motion are found in many biological contexts such as actin filaments, molecular
motors walking on microtubules, flagellar bundles and ciliary hairs. These
cellular components maintain structure, influence signaling, and provide motile
forces. Much recent effort has been directed at synthesising biomimetic
analogues of such components. Motor-microtubule mixtures \cite{nedelec1997} and
self-assembled motor-microtubules bundles capable of motility
\cite{sanchez2012} are two examples where dynamic structures remarkably similar
to those occurring \emph{ in vivo} have been synthesized in the laboratory.
These emergent structures  are maintained by nonequilibrium ``active'' forces
generated by the consumption of chemical energy. Hydrodynamic flow provides a
way of transmitting nonequilibrium forces and its role in creating and
maintaining dynamic structures remains relatively unexplored.

Our present theoretical understanding of hydrodynamics in nonequilibrium
energy-converting  systems is based largely on continuum  \cite{marchetti2013}
or kinetic \cite{saintillan2013active} theories that prescribe, respectively,
the spatiotemporal evolution of coarse-grained fields or distribution
functions. These theories predict long wavelength collective phenomena like
instabilities \cite{simha2002a, saintillan2008a}, the onset of flows
\cite{woodhouse2012spontaneous} through spontaneous symmetry breaking
\cite{tjhung2012spontaneous} and  the continuous generation of topological
defects \cite{giomi2013defect}. The complementary domain of short wavelength
phenomena where near-field hydrodynamics, the shape of the active body and
excluded volume interactions are important, has been much less studied
\cite{jayaraman2012, *laskar2013, chelakkot2014flagellar, *jiang2014motion}.
This domain is relevant in situations where particle numbers are not
thermodynamically large and a hydrodynamic limit is not necessarily attained,
as for example in the cell. Theoretical descriptions that do not presume the
existence of continuum and hydrodynamic limits, then, are required to provide
insight into the physics of short wavelength phenomena in active matter.  

In this Letter we study short wavelength aggregation phenomena in suspensions
of stiff, apolar, active filaments using a description that resolves individual
filaments and includes their hydrodynamic and excluded volume interactions. Our
model for an active filament consists of rigidly connected active beads that
produce force-free, torque-free flows in the fluid \cite{jayaraman2012}. We
approximate the active flow of each bead by its slowest decaying component, the
irreducible stresslet \cite{ghose2014}. The stresslet principal axis is always
aligned along the filament tangent and the principal values are either positive
(extensile) or negative (contractile). This ensures that an isolated straight
filament produces active flows but does not actively translate or rotate and
is, thus, individually apolar. However, as we show, translational and
rotational motion is possible in the \emph{combined} flow of two or more stiff,
apolar filaments. This inherently collective and nonequilibrium source of
motion, when constrained by the anisotropic excluded volume interactions
between filaments, leads to non-trivial states of aggregation in suspension. 

The apolar filament suspension is an example of a system where forces that
break time reversal invariance, and which are usually associated with
dissipation and the cessation of motion, drive the system and lead to
nonquilibrium steady states (NESS) with non-zero fluxes.  A well-known paradigm
of NESS is the driven dissipative system, where external forces at the system
boundary produce nonequlibrium. In contrast, here internal forces distributed
continuously within the system boundary produce nonequilibrium.  Therefore,
apolar active suspensions are novel model systems for studying NESS both
theoretically and experimentally. Many recent developments in the statistical
physics of NESS, such as fluctuation theorems, macroscopic fluctuation theory,
and motility-induced phase separation can be tested in the system we study
here. 

\begin{figure*}[t] \includegraphics[width=1.02\textwidth]{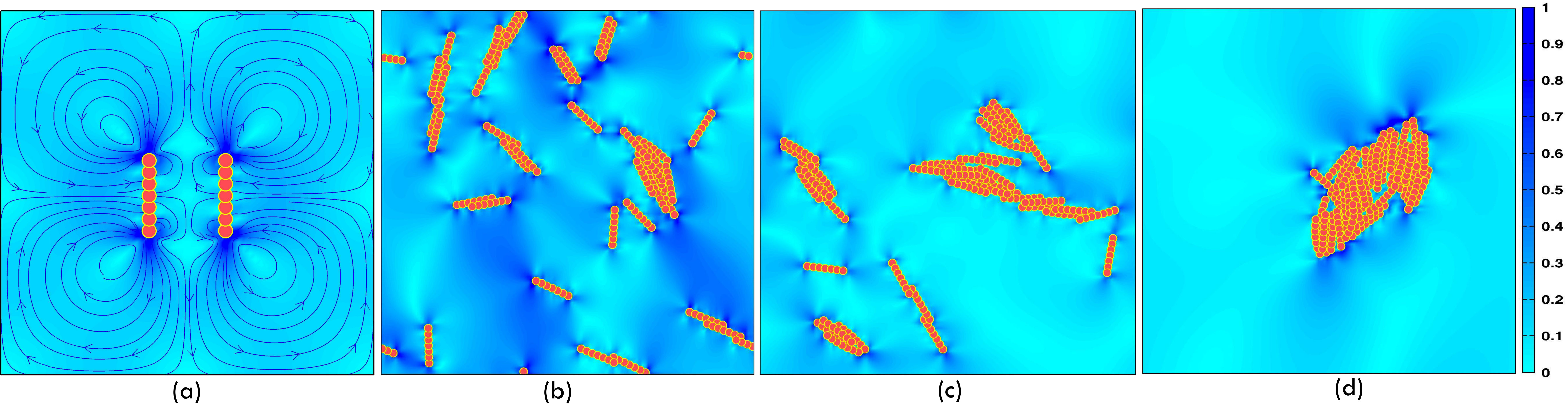}
\caption{\label{kinetics}(Color online) Self-assembly of extensile filaments.
(a) Flow field of two extensile filaments producing lateral attraction. (b)
Lateral attraction drive rapid dimerization at early times. (c) Coagulation of
$k$-mers at intermediate times. (d) Aggregation of $k$-mers into a single
cluster at late times. The  background color indicates the magnitude of the
velocity normalized by its maximum. Snapshot b-d are for a $\phi = 0.05$ system taken at $t = 0.1\tau_e$, $t = 0.2\tau_e$ and $t = 1.8\tau_e$ respectively.
} \label{fig1} \end{figure*}

Our results are as follows. We find distinct fixed points for the non-linear
dynamics of pairs of filaments translating and rotating in their mutual flow.
Extensile filaments, driven by dominant lateral attractions, join into stable
rotating dimers. Contractile filaments, driven by dominant perpendicular
attractions, join into translating and rotating T-shaped dimers. Extensile
filament suspensions are always unstable and show a two-step aggregation
kinetics.  A rapid initial growth of small clusters is followed by a slower
merging of these clusters into large, dynamic aggregates which translate and
rotate ballistically.  Contractile filament suspensions remain homogeneous and
isotropic but develop short wavelength microstructure that depends on volume
fraction.  We observe such microstructures in a sequence of states of asters,
clusters, and incipient gels with increasing volume fraction. The centres of
mass of the filaments show a non-thermal diffusion which decreases with
increasing volume fraction. We describe these results in detail below and then
discuss their significance. 

\emph{Active Zimm model:}  We briefly describe our model for an active filament
\cite{jayaraman2012}. We consider $N_{b}$ spherical active beads of radius $a$
centred  at ${\bf r}_n$. These are acted on by elastic forces ${\bf F}_n =
-\partial U({\bf r}_1, \ldots, {\bf r}_{N_b})/\partial{\bf r}_n$ derived from
the potential $U({\bf r}_1, \ldots, {\bf r}_{N_b})$ which bonds neighbouring
beads into a connected filament, penalises filament curvature, and enforces
volume exclusion between beads. The equilibrium bond length is $b$. Each bead
has a stresslet ${\bf S}_n = S_0({\bf t}_n{\bf t}_n - \frac13{\mathbf{I}})$
aligned along the local tangent ${\bf t}_n$. The beads move in response to the
elastic forces acting on them and the hydrodynamic flow which results, at low
Reynolds numbers and small bead size,  in the following equation of motion 
\begin{equation} \dot{\bf r}_n = \mu {\bf F}_n + \sum_{m \neq n}\left[ {\mathbf
G}({\bf r}_n - {\bf r}_m) \cdot {\bf F}_m + \mathbf{\nabla}{\mathbf G}({\bf
r}_n - {\bf r}_m)\cdot{\bf S}_m\right].  \end{equation}
Here, ${\mathbf G}({\bf r})$ is a Green function of the Stokes equation, $\mu$
is the bead mobility and $S_0$ is the principal value of the stresslet. The
flow produced by a single bead is extensile or contractile when $S_0$ is,
respectively, positive or negative. The stresslet self-term is absent as it
cannot produce translational motion in an unbounded domain. These equations of
motions may be thought of as a generalization, to active filaments, of the Zimm
model of polymer dynamics where hydrodynamic interactions are treated at the
level of the Kirkwood-Risemann superposition approximation. The accuracy of
this approximation is $(a/b)^3$ for monopolar flows and, by an extension of the
original study, $(a/b)^4$ for dipolar flows. With the choice of parameters made
here, we make no more than 6$\%$ error in computing hydrodynamic interactions
using the superposition approximation.  We simulate a suspension of $N$ such
filaments using the lattice Boltzmann method to compute the active flows, as
described in detail in \cite{jayaraman2012}.

\begin{figure}[b]  
 \centering
\includegraphics[width=0.45\textwidth]{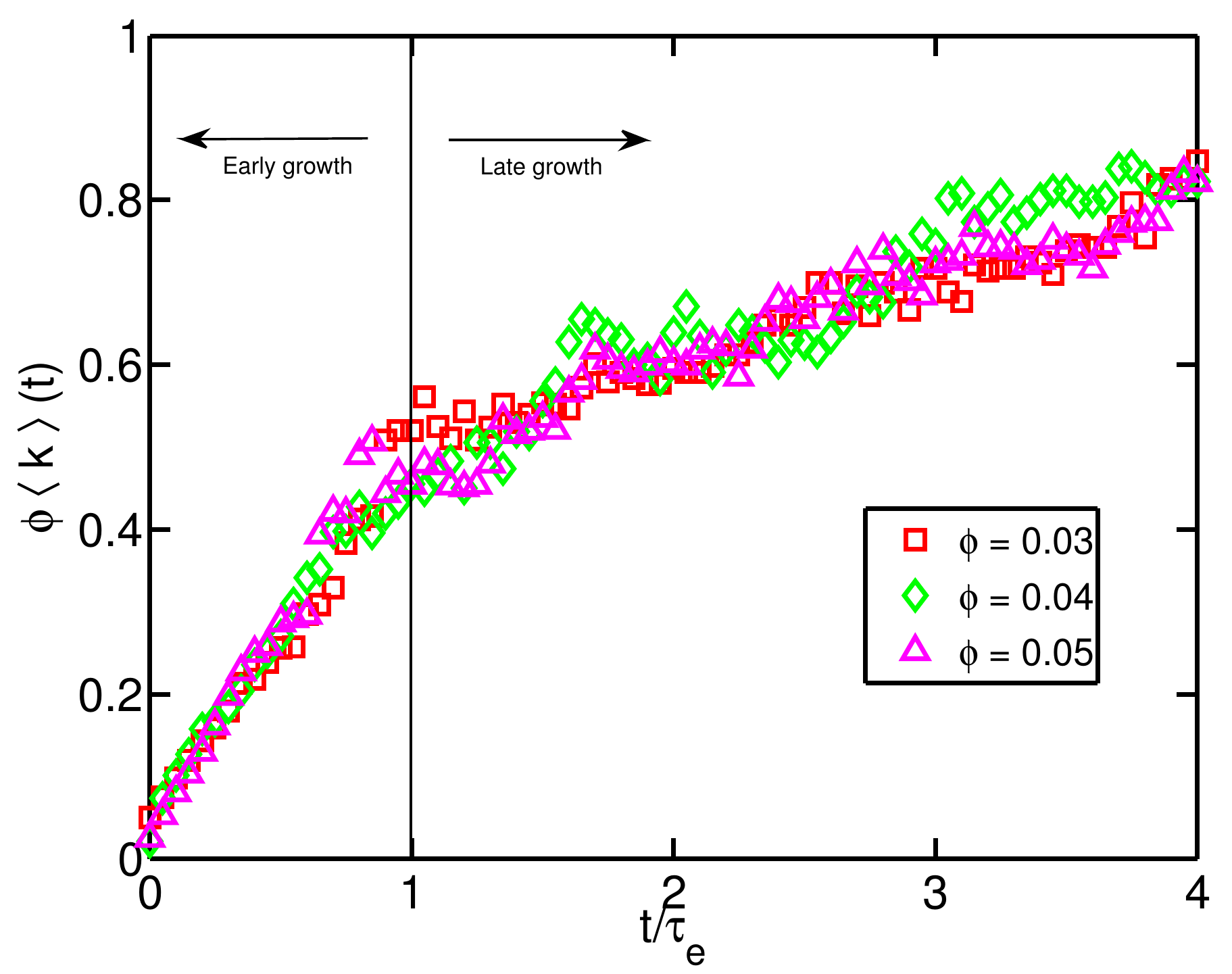} 
\caption{(Color online) \label{fig:meancluster_ext} Growth kinetics in extensile filaments.  The mean of the $k$-mer distribution function $\langle k \rangle$    as a function of time $t$, for varying filament numbers $N$. Dimerization dominates the rapid growth for $t \ll \tau_e$ while coagulation of $k$-mers dominates the slow growth  for $t \gg \tau_e$. These regimes correspond, respectively, to panels (b) and (c) of  Fig.(\ref{fig1}).}
\label{fig2}
\end{figure}
\begin{figure*}[t]
\includegraphics[width=1.02\textwidth]{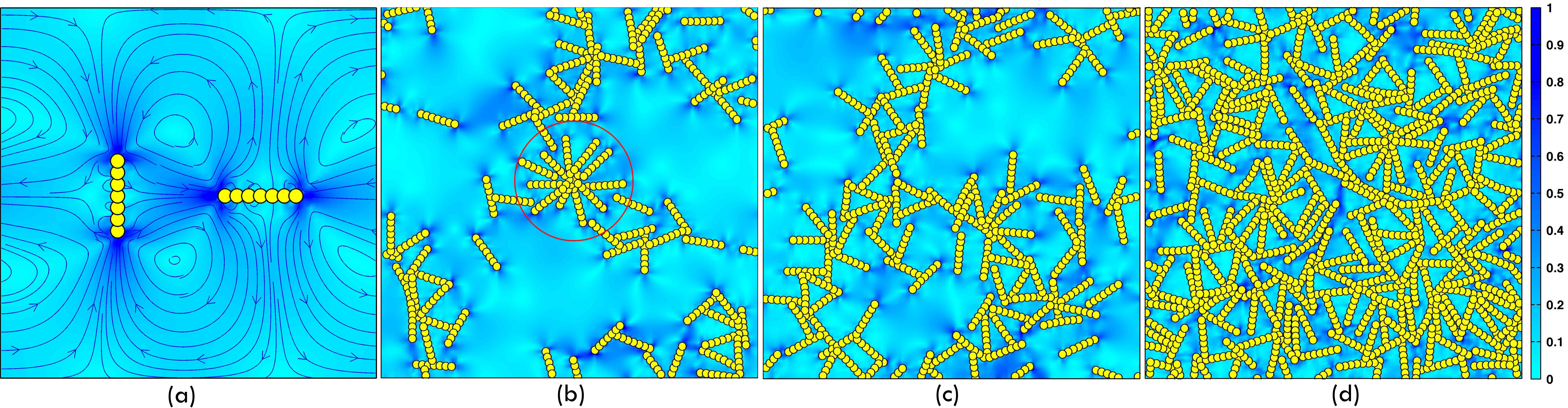}
\caption{\label{aster} (Color online) Self-assembly of contractile filaments. (a) Flow field of two contractile filaments producing perpendicular attraction. Microstructured steady states of asters at volume fraction $\phi = 0.11$ in (b),  clusters at $\phi = 0.15$ in (c)  and incipient gel at $\phi = 0.34$ in (d). A typical aster configuration is circled in (b). The background color indicates the magnitude of the velocity normalized by its maximum.}
\label{fig3}
\end{figure*}

\emph{Activity induced dimerization :} As a prelude to collective dynamics we study the motion of a pair of active filaments in their mutual hydrodynamic flow. The flow points outwards along the long-axis for extensile filaments but inwards for contractile filaments. More generally, contractile flow is obtained from extensile flow by reversing all streamlines \cite{jayaraman2012}. Extensile filament pairs \emph{attract laterally} as is clear from the streamlines around a proximate pair shown in Fig.(\ref{fig1}a). Extensile filament pairs \emph{repel perpendicularly} as is clear from the \emph{reversed} streamlines of Fig.(\ref{fig3}a).  The lateral attraction drives parallel filaments towards each other till steric forces prevent further motion. The shear flow along the filament long axis causes a small lateral motion, thus forming a stable rotating dimer (see movie in \cite{SI}). This is the stable state for a pair of extensile filaments and is  independent of the special initial condition chosen in the simulation.  Streamlines around a  contractile filament flow in a direction opposite to that of extensile filaments.  Thus, contractile filament pairs \emph{repel laterally} and \emph{attract perpendicularly} in their mutual flow as is  clear from the streamlines in Fig.(\ref{fig1}a) and Fig.(\ref{fig3}a). The perpendicular attraction, combined with steric repulsion,  drives contractile filaments to form $T$-shaped dimers which translate and rotate (see movie in \cite{SI}). Again, this stable state is independent of the initial condition used to generate it. 

The dimerization described above is an essentially nonequilibrium phenomenon mediated by the active hydrodynamic flow. Although the microscopic processes that give rise to activity-induced flows and result in the dimerization phenomenon might vary greatly in origin and function, the resultant hydrodynamic flows and interactions can be completely parametrised using the irreducible expansion of the active flow \cite{ghose2014}, the leading term of which is retained in this analysis. This makes the nonequilibrium dimerisation both robust and generic. 

\emph{Aggregation of extensile filaments :}  The nonequilibrium flow-induced forces described above produce self-assembled structures in suspensions of filaments which depend crucially on the sign of the stresslet. In extensile filaments, the lateral hydrodynamic attraction destabilises the homogeneous isotropic state of the suspension. Dimer pairs form rapidly and further aggregate to form $k$-mers. Asymptotically,  the individual clusters collapse into a single dynamic cluster  which translates and/or rotates according to its final shape. The qualitative features of the  aggregation kinetics is shown in the panels (b-d) of Fig.\ref{fig1} and in the movie in \cite{SI}.

To quantitatively understand the aggregation kinetics we compute the distribution $P(k, t)$  of $k$-mers, which we define as $k$ filaments that are within half a filament length of each other. The distribution is computed over an ensemble of $50$ simulations with different initial conditions. The first moment 
$\langle k \rangle(t) = \sum_{k=1}^{\infty} k\, P(k, t)$  provides a measure of the mean size of the aggregates and this quantity is plotted in Fig.(\ref{fig2}). There is a rapid initial growth of the mean  size for a duration $\tau_e$ followed by a slower growth at later times. We estimate $\tau_e$ from the intersection of a pair of straight lines that best fit the data. From this, we  find that the duration $\tau_e$ of rapid growth is weakly dependent or, possibly, independent of volume fraction while the  duration of slow growth to saturation increases  with volume fraction. 

The nonequilibrium aggregation proceeds through a frequent coagulation process in which $k$-mers join  $j$-mers to form $(k+j)$-mers and a rare fragmentation process in which $(k+j)$-mers break into $j$-mers and $k$-mers. A careful observation of the kinetics reveals a hierarchy of rates: dimerization is the most rapid process, followed by single filaments joining $k$-mers (with $k \geq 3$), then dimers joining $k$-mers and, finally,  $k$-mers joining $j$-mers (with $j \geq 3$). The fragmentation rates show an opposite hierarchy: the breaking  of a $(k+j)$-mer into $k$-mers and $j$-mers with $k \approx j$ is the most rapid process, followed the breaking of dimers and, finally, monomers from $(k+j)$-mers. These rates are determined by the hydrodynamic flow due to $k$-mers and $j$-mers and is sensitive to their shape. A direct calculation of the coagulation and fragmentation rates is, therefore, a formidable problem.  Nonetheless, constructing a tractable theoretical model for the rates of  hydrodynamic coagulation and fragmentation is desirable as the rates can then be employed in a Smoluchowski-like mean-field description of aggregation kinetics. 
\begin{figure}[t]  
\subfigure[~Stationary cluster size distribution]{
 \includegraphics[width=0.45\textwidth]{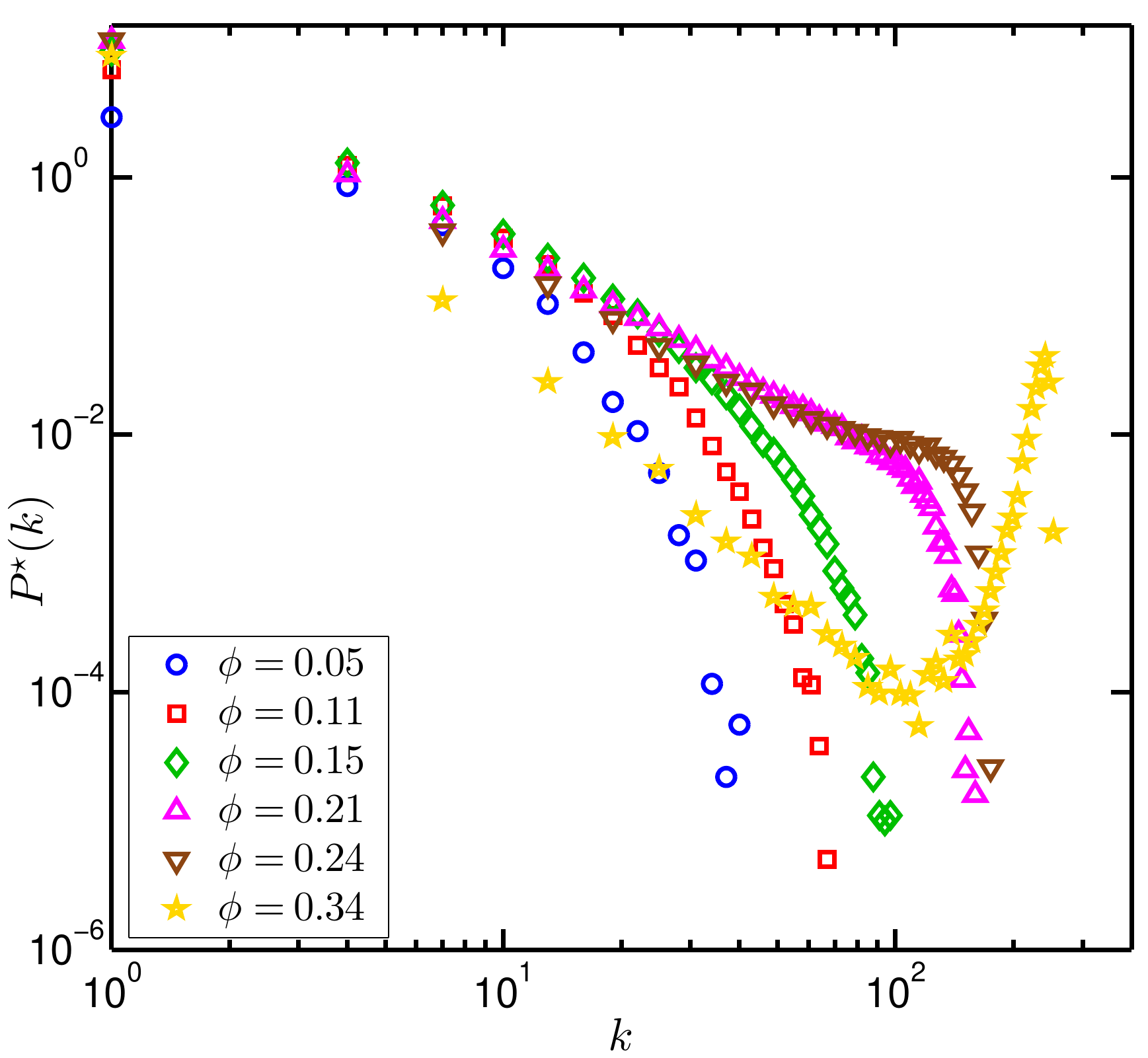}} 
\subfigure[~Filament center of mass diffusion]{
\includegraphics[width=0.45\textwidth]{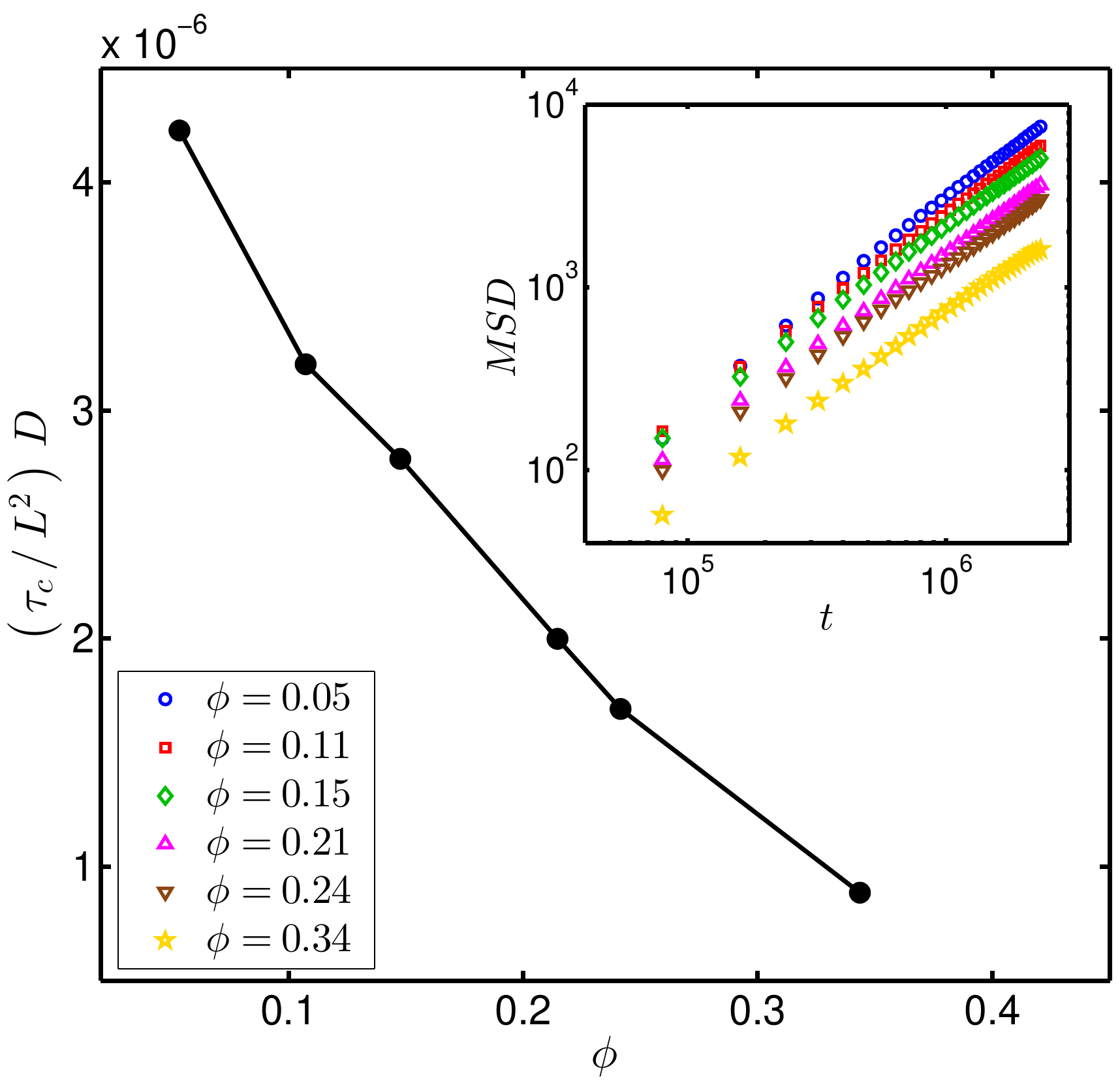} }
\caption{(Color online) \label{fig:con_msd} (a) Steady state $k$-mer distribution function $P^{\star}(k)$ with varying volume fraction $\phi$. The distribution shows two qualitative changes at $\phi = 0.15$ and $\phi = 0.34$. (b) Filament diffusion coefficient $D$ as a function of volume fraction $\phi$ and mean square displacement (MSD) as a function of time $t$ (inset).  $D$ is non-dimensionalised by $L^2/\tau_c$, where $L = (N_b-1)b$ is the filament length and $\tau_c$ the typical time taken to reach stationarity from initially ordered states.
}
\label{fig4}
\end{figure}

\emph{Microstructure of  contractile filaments:} Nonequilibrium self-assembly in contractile filaments proceeds in a manner very different from extensile filaments. Here, perpendicular hydrodynamic attraction induces dynamic microstructure in initially homogeneous and isotropic states of the suspension. Averaged over long times and large lengths, the suspension remains homogeneous and isotropic, but any instantaneous configuration reveals both positional inhomogeneities and orientational anisotropies. Three distinct microstructures emerge as a function of volume fraction, as shown in panels (b-d) of Fig.(\ref{fig3}). At the lowest volume fractions, filaments transiently cluster into star-shaped $k$-mers reminscent of asters (panel b). These $k$-mers frequently break and reform while their centers of mass diffuse as shown in the movie in \cite{SI}. Through this dynamics, the spatiotemporally averaged position and orientation become, respectively, uniform and isotropic. With an increase in volume fraction, the steric constraints limit the dynamic joining and breaking of $k$-mers and their diffusion (panel c). A percolating network of filaments is formed which, at short time scales, may present a solid-like response by being able to support forces and torques. With further increase in volume fraction, this network gets sterically constrained to a great extent and thus shows greatly reduced positional and orientational dynamics. States that are initially inhomogeneous and have a preferred orientation are  unstable and evolve, rapidly, into homogeneous, isotropic microstructured states shown in the movie \cite{SI}.

In order to distinguish between the distinct microstructural states, we compute the stationary $k$-mer distribution $P^{\star}(k) = \lim_{t\rightarrow\infty}P(k, t)$ for different filament volume fractions and plot it in Fig. (\ref{fig4}a). Asters correspond to $\phi=0.05, 0.11, 0.15$, clusters to $\phi = 0.21, 0.24$ and the incipient gel to $\phi = 0.34$. The distribution shows qualitatively different shapes for each microstructure and the rapid change in shape around $\phi = 0.3$ provides a signature of gelation. The motion of individual filaments is diffusive in all three microstructured states as the  linear variation of the mean square displacement with time shows in the inset of Fig.(\ref{fig4}b). The diffusive constant $D$ decreases with volume fraction as Fig.(\ref{fig4}b) shows. The parallel hydrodynamic repulsion between filaments enhances their effective volume which increases the steric hindrance and so constrains filament motion, leading to smaller diffusion. It is important to emphasise that the diffusive motion occurs in the absence of thermal fluctuations and consequently $D$ does not obey a Stokes-Einstein relation. 

\emph{Discussion:} Hierarchical self-assembly in thermodynamic equilibrium is driven by a competition between energy and entropy. In contrast, the self-assembly mechanism reported here is an essentially nonequilibrium phenomenon driven by  dissipative, hydrodynamic flow produced by active processes. The relative translation of pairs of particles, which contributes to microstructure formation, is possible to due the absence of Onsager symmetry of the mobility matrix that prevents such motion in passive hydrodynamic flow.  This short wavelength physics is resolved by our particle picture of active flow but is beyond the reach of current long wavelength continuum and kinetic theories. 

Several recent experiments report dynamic self-assembly of anisotropic particles in hydrodynamic flow. The formation of  ``living'' colloidal crystals on planar substrates \cite{palacci2013living} bears a striking similarity with the aggregation of extensile filaments into liquid crystalline clusters. We suggest that extensile, dipolar hydrodynamic flows, modified by the presence of a no-slip boundary to induce attraction, play a crucial role in the self-assembly of such ``living'' crystals. The dimerization of polar bi-metallic catalytic nanomotors \cite{mallouk2013} is, again, strikingly similar to the dimerization we report here.  To directly verify some of our predictions we suggest experiments on  apolar active particles made by adding a third metallic stripe of gold or platinum to bi-metallic catalytic nanomotors.  Finally,  we note that  our work suggest a new nonequilibrium mechanism with which to control and manipulate hierarchical self-assembly of soft matter.

Computing resources through HPCE, IIT Madras and Annapurna, IMSc are gratefully acknowledged. RA thanks the Indo-US Science and Technology Foundation for a fellowship. The authors thank M. E. Cates, P. Chaikin, S. Ghose,  A. Laskar, T. E. Mallouk, H. Masoud, D. Pine, M. Shelley and R. Singh for helpful discussions, S. Ghose for help with figures and R. Manna for sharing simulation data on filament pairs.

\end{document}